\documentclass[twocolumn,amsmath,aps,fleqn]{revtex4}
\usepackage{graphicx,amssymb}
\begin{document}
\newcommand{\be}{\begin{equation}}
\newcommand{\ee}{\end{equation}}
\newcommand{\bq}{\begin{eqnarray}}
\newcommand{\eq}{\end{eqnarray}}
\newcommand{\bsq}{\begin{subequations}}
\newcommand{\esq}{\end{subequations}}
\newcommand{\bc}{\begin{center}}
\newcommand{\ec}{\end{center}}
\newcommand {\R}{{\mathcal R}}
\newcommand{\al}{\alpha}
\newcommand\lsim{\mathrel{\rlap{\lower4pt\hbox{\hskip1pt$\sim$}}
    \raise1pt\hbox{$<$}}}
\newcommand\gsim{\mathrel{\rlap{\lower4pt\hbox{\hskip1pt$\sim$}}
    \raise1pt\hbox{$>$}}}

\title{Perfect fluid Lagrangian and its cosmological implications in theories of gravity with nonminimally coupled matter fields}

\author{P.P. Avelino}
\email[Electronic address: ]{pedro.avelino@astro.up.pt}
\affiliation{Instituto de Astrof\'{\i}sica e Ci\^encias do Espa{\c c}o, Universidade do Porto, CAUP, Rua das Estrelas, PT4150-762 Porto, Portugal}
\affiliation{Centro de Astrof\'{\i}sica da Universidade do Porto, Rua das Estrelas, PT4150-762 Porto, Portugal}
\affiliation{Departamento de F\'{\i}sica e Astronomia, Faculdade de Ci\^encias, Universidade do Porto, Rua do Campo Alegre 687, PT4169-007 Porto, Portugal}

\author{R.P.L. Azevedo}
\email[Electronic address: ]{rplazevedo@fc.up.pt}
\affiliation{Centro de F\'isica do Porto}
\affiliation{Departamento de F\'{\i}sica e Astronomia, Faculdade de Ci\^encias, Universidade do Porto, Rua do Campo Alegre 687, PT4169-007 Porto, Portugal}

\date{\today}
\begin{abstract}

In this paper we show that the on-shell Lagrangian of a perfect fluid depends on microscopic properties of the fluid, giving specific examples of perfect fluids with different on-shell Lagrangians but with the same energy-momentum tensor.  We demonstrate that if the fluid is constituted by localized concentrations of energy with fixed rest mass and structure (solitons) then the average on-shell Lagrangian of a perfect fluid is given by ${\mathcal L}_m=T$, where $T$ is the trace of the energy-momentum tensor. We show that our results have profound implications for theories of gravity where the matter Lagrangian appears explicitly in the equations of motion of the gravitational and matter fields, potentially leading to observable deviations from a nearly perfect cosmic microwave background black body spectrum: $n$-type spectral distortions, affecting the normalization of the spectral energy density. Finally, we put stringent constraints on $f(R,{\mathcal L}_m)$ theories of gravity using the COBE-FIRAS measurement of the spectral radiance of the cosmic microwave background.

\end{abstract}
\maketitle

\section{\label{intr}Introduction}

In some theories of gravity the matter Lagrangian enters explicitly in the equations of motion of the gravitational and matter fields, thus making the quest for the appropriate form of the Lagrangian a relevant one, in particular in the case where the energy-momentum tensor of the matter fields is described by a perfect fluid. In \cite{Brown:1992kc}, by choosing a specific form of the action for a perfect fluid, it has been shown that, in the context of general relativity (GR), the on-shell Lagrangian of a perfect fluid is equal to ${\mathcal L}_m = p$, where $p$ is the proper pressure. There it has also been noted that the addition of appropriate surface integrals to the action would lead to ${\mathcal L}_m = -\rho$, where $\rho$ is the proper energy density, without affecting the equations of motion. Although this degeneracy has no observable consequences in the context of GR,  in theories of gravity where the matter Lagrangian appears explicitly in the equations of motion of the gravitational and matter fields, such as $f(R,{\mathcal L}_m)$ \cite{Nojiri:2004bi,Allemandi:2005qs,Bertolami:2007gv,Sotiriou:2008it,Harko:2010mv} and $f(R,T)$ \cite{2011PhRvD..84b4020H} theories of gravity, the dynamics of the gravitational and matter fields may be significantly different depending on whether one considers ${\mathcal L}_m = p$ or ${\mathcal L}_m = -\rho$ as the Lagrangian of a perfect fluid \cite{Faraoni:2009rk}. Nevertheless, in \cite{Bertolami:2008ab} it has been argued that ${\mathcal L}_m = -\rho$ is the correct choice for non-minimally coupled (NMC) theories, a suggestion that has been followed in some of the subsequent works investigating astrophysical and cosmological consequences of such theories (see, e.g.  \cite{Nesseris:2008mq,Bertolami:2009ic,Azizi:2014qsa,Ribeiro:2014sla,Azevedo:2016ehy}).

In \cite{Minazzoli:2012md} (see also \cite{Harko:2010zi}) the Lagrangian of a barotropic perfect fluid whose proper pressure is a function solely of the rest mass density was derived from the equations of motion, with minimal assumptions about the gravitational theory --- it was assumed that the matter Lagrangian is independent of the derivatives of the metric and that the particle number is conserved. Still, even in the simplest cases relevant to cosmology, the proper pressure depends not only on the rest mass density (i.e. the proper number density times the rest mass of the particles), but also on the root mean square velocity of the particles. Moreover, the assumed dependence of the Lagrangian on the rest mass density, rather than the energy density, limits the application of the result presented in \cite{Harko:2010zi} to the observed content of our Universe, even during the radiation era.

In the present paper we revisit the problem of finding the on-shell Lagrangian of a perfect fluid, focusing on its dependence on the microscopic properties of the fluid. We start, in Sec. II, by briefly reviewing the equations of motion of the gravitational and matter fields in the context of $f(R,{\mathcal L}_m)$ gravity. In Sec. III we explicitly demonstrate that the on-shell Lagrangian of a perfect fluid depends on the microscopic properties of the fluid not specified by its energy-momentum tensor. By modeling particles as topological solitons in 1+1 dimensions, we also determine the (averaged) on-shell Lagrangian of a perfect fluid composed of such particles, and we investigate the corresponding dynamics in flat 1+1 dimensional Friedmann-Robertson-Walker (FRW) universes in the context of $f(R,{\mathcal L}_m)$ theories of gravity. In Sec. IV we extend the results of the previous section to 3+1 dimensions by considering a perfect fluid constituted by particles which can be described as localized concentrations of energy with fixed rest mass and structure (solitons), studying the evolution of their linear momentum in a homogeneous and isotropic Friedmann-Robertson-Walker universe. In Sec. V we apply the results of Sec. IV to photons, and constrain the spectral distortions of the cosmic microwave background induced in $f(R,{\mathcal L}_m)$ gravity, using the COBE-FIRAS measurement of the spectral radiance of the cosmic microwave background.

Throughout this paper we use units such that $c=1$, where $c$ is the value of the speed of light in vacuum, and we adopt the metric signature $(-,+,+,+)$. The Einstein summation convention will be used when a greek index variable appears twice in a single term, once in an upper (superscript) and once in a lower (subscript) position.

\section{$f(R,{\mathcal L}_m)$ gravity}

Consider the action
\be
S=\int f(R,{\mathcal L}_m) {\sqrt {-g}}d^4 x \,, \label{action}
\ee
allowing for NMC matter fields. Here, ${\mathcal L}_m$ is the matter Lagrangian, $R$ is the Ricci scalar, $g=\det (g_{\mu\nu})$ and $g_{\mu\nu}$ are the components of the metric tensor. The equations of motion of the gravitational field,
\be
f_{,R} G_{\mu \nu} = \frac{1}{2}g_{\mu \nu}(f-Rf_{,R})+\Delta_{\mu \nu} f_{,R} +\frac{1}{2} f_{, {\mathcal L}_m} (T_{\mu \nu} -{\mathcal L}_m g_{\mu \nu})\,,\label{grav}
\ee
may be obtained by minimizing the action with respect to variations of the metric. Here, a comma denotes a partial derivative,  $\nabla_\mu$ represents a covariant derivative with respect to $x^{\mu}$, $\Delta_{\mu \nu} \equiv \nabla_\mu \nabla_\nu - g_{\mu \nu} \Box$, $\Box \equiv \nabla^\mu \nabla_\mu$,
\be
G_{\mu \nu}\equiv R_{\mu \nu} - \frac{1}{2} g_{\mu \nu} R\,,
\ee
and 
\be
T^{\mu\nu}=-\frac{2}{{\sqrt {-g}}} \frac{\delta({\mathcal L}_m{\sqrt {-g}})}{\delta g_{\mu \nu}}=-2\frac{\delta{\mathcal L}_m}{\delta g_{\mu \nu}}+g^{\mu\nu} {\mathcal L}_m \label{T}
\ee
are the components of the energy momentum-tensor of the matter fields. Taking into account that $ \nabla_\mu G^{\mu \nu}=0$, it is simple to show that the energy-momentum tensor is not in general covariantly conserved:
\be
\nabla_{\mu} T^{\mu \nu} =  S^{\nu}\,, \label{Tcons}
\ee
where
\bq
S^{\nu} &=& ({\mathcal L}_m g^{\mu \nu} -T^{\mu \nu}) \times \nonumber \\
&\times& \left( [\ln \left| f_{,{\mathcal L}_m}\right|]_{,R} \nabla_\mu R + [\ln \left|f_{,{\mathcal L}_m}\right|]_{,{\mathcal L}_m} \nabla_\mu {\mathcal L}_m  \right)\,.\label{Tcons1}
\eq
On the other hand, in $3+1$ dimensions the trace of Eq. (\ref{grav}) is given by
\be
\Box f_{,R} = \frac16 \left[4f-2R f_{,R}+f_{,{\mathcal L}_m}(T-4{\mathcal L}_m)\right]\,.  \label {boxR}
\ee
Equations (\ref{grav}), (\ref{Tcons}) and  (\ref{boxR}) show that in $f(R,{\mathcal L}_m)$ gravity ${\mathcal L}_m$ appears explicitly in the equations of motion of both the gravitational and the matter fields, thus implying that the knowledge of the Lagrangian of the matter fields is essential in order to obtain the corresponding dynamics.

\section{Scalar matter fields}

Assume, for the moment, that the matter fields are described by a real scalar field $\phi$ governed by the a generic Lagrangian of the form ${\mathcal L}_m(\phi,X)$, where
\begin{equation}\label{eq:kinetic_scalar1}
X=-\frac{1}{2}\nabla^\mu \phi \nabla_\mu \phi 
\end{equation}
is the kinetic term. The Euler-Lagrange equation for the scalar field $\phi$ may be obtained by minimizing the action with respect to variations of $\phi$, and is given by 
\be
0 = - \frac{\partial f}{\partial  \phi} + \nabla_\mu \left[ \frac{\partial f }{\partial (\nabla_\mu \phi)}\right]\,, \label{eqofmpsi}
\ee
On the other hand, Eq. (\ref{T}) implies that
\be
\label{eq:fluid}
T^{\mu\nu} ={\mathcal L}_{m,X} \nabla^\mu \phi \nabla^\nu \phi + {\mathcal L}_m g^{\mu \nu}\,.
\ee

\subsection{Perfect fluid with ${\mathcal L}_m=p$}

For timelike $\nabla_\mu \phi$, it is possible to write the energy-momentum tensor in a perfect fluid form
\begin{equation}\label{eq:fluid2}
T^{\mu\nu} = (\rho + p) u^\mu u^\nu + p g^{\mu\nu} \,,
\end{equation}
by means of the following identifications
\begin{equation}\label{eq:new_identifications}
u_\mu = \frac{\nabla_\mu \phi}{\sqrt{2X}} \,,  \quad \rho = 2 X p_{,X} - p \, ,\quad p =  {\mathcal L}_m(\phi,X)\, .
\end{equation}
In Eq.~(\ref {eq:fluid2}), $u^\mu$ are the components of the 4-velocity field describing the motion of the fluid, while $\rho$ and $p$ are its proper energy density and pressure, respectively. Observe that, in this case, ${\mathcal L}_m=p$ which in the context of GR is one of the possible choices considered in the literature for the on-shell Lagrangian of a perfect fluid. Note that, since the 4-velocity is a timelike vector, the correspondence between scalar field models of the form  ${\mathcal L}_m={\mathcal L}_m (\phi,X)$ and perfect fluids breaks down whenever $\nabla_\mu \phi$ is spacelike, as is the case of non-trivial static solutions. In the case of a homogeneous and isotropic universe filled with a perfect fluid with arbitrary density $\rho$ and $p=0$, Eqs. (\ref{eq:fluid2}) and  (\ref{eq:new_identifications}) imply that the dynamics of this fluid may be described by a matter Lagrangian whose on-shell value is equal to zero everywhere (a simple realization of this situation would be to take ${\mathcal L}_m=X-V= {\dot \phi}^2/2-V(\phi)$ with the appropriate potential $V$ and initial conditions, so that $V(\phi)$ is always equal to ${\dot \phi}^2/2$ --- here a dot represents a derivative with respect to the physical time).

\subsection{Scalar particles in 1+1 dimensions: ${\mathcal L}_m=T$}

Throughout most of its history the energy content of the Universe is expected to have been dominated by moving particles that may be modeled as localized concentrations of energy with fixed rest mass and structure. In this section, our particle shall be modeled as a topological soliton of the field $\phi$ in 1+1 dimensions. For concreteness, assume that the matter fields may be described by a real scalar field with Lagrangian 
\be
{\mathcal L}_m= -\frac12 \partial_\mu \phi \partial^\mu \phi  - V(\phi)\,,
\ee
where $V(\phi) \ge 0$ is a real scalar field potential
\be
V(\phi)=\frac{\lambda}{4} \left(\phi^2-\eta^2\right)^2\,,
\ee
which has two degenerate minima at $\phi=\pm \eta$. Further on, we shall demonstrate that the main results derived in this paper do not rely on this specific choice for the Lagrangian of the matter fields.

In this case, ${\mathcal L}_{m,X}=1$ and the energy-momentum tensor of the matter fields is given by
\be
\label{eq:fluid1}
T^{\mu\nu} = \nabla^\mu \phi \nabla^\nu \phi + {\mathcal L}_m g^{\mu \nu}\,.
\ee
On the other hand, the equation of motion for the scalar field $\phi$ is 
\bq
\Box \phi&=&-\left( [\ln \left| f_{,{\mathcal L}_m}\right|]_{,R} \nabla_\mu R \right. \nonumber\\
&+&\left. [\ln \left|f_{,{\mathcal L}_m}\right|]_{,{\mathcal L}_m} \nabla_\mu {\mathcal L}_m \right)\nabla^\mu \phi +V_{,\phi}\,. \label{dalphi}
\eq
Multiplying Eq. (\ref{dalphi}) by $\nabla^\nu \phi$, and taking into account that $T^{\mu\nu} - {\mathcal L}_m g^{\mu \nu}= \nabla^\mu \phi \nabla^\nu \phi$, one recovers Eq. (\ref{Tcons}).

\subsubsection{Minkowski spacetime}

In a $1+1$ dimensional Minkowski space-time the line element can be written as $ds^2=-dt^2+dz^2$.  Hence, neglecting the self-induced gravitational field, the Lagrangian and the equation of motion of the scalar field $\phi$ are given respectively by
\bq
{\mathcal L}_m&=& \frac{{\dot \phi}^2}{2} -\frac{\phi'^2}{2}-V(\phi)\,, \label{Lagphi}\\ 
{\ddot \phi} - \phi''&=& -\frac{dV}{d \phi}\,, \label{phieqmM}
\eq
where a dot denotes a derivative with respect to the physical time $t$ and a prime represents a derivative with respect to the spatial coordinate $z$.

The components of the energy-momentum tensor of the particle can now be written as
\bq
\rho_\phi=-{T^{0}}_0&=&\frac{{\dot \phi}^2}{2}+\frac{\phi'^2}{2}+V(\phi)\,,\\
T^{0z}&=&-{\dot \phi}\phi'\,,\\
p_\phi={T^{z}}_z&=&\frac{{\dot \phi}^2}{2}+\frac{\phi'^2}{2}-V(\phi)\,,
\eq
so that the trace $T$ of the energy-momentum tensor is given by
\be
T={T^\mu}_\mu= {T^{0}}_0+{T^{z}}_z =-\rho_\phi+p_\phi=-2 V(\phi)\,. \label{Ttrace}
\ee

Consider a static soliton with $\phi=\phi(z)$. In this case Eq. (\ref{phieqmM}) becomes
\be
 \phi''= \frac{dV}{d\phi} \label{phieqmM1}\,,
\ee
and it can be integrated to give
\be
\frac{\phi'^2}{2} = V\,,\label{KeqU}
\ee
assuming that $|\phi| \to \eta$ for $z \to \pm \infty$. If the particle is located at $z=0$,  Eq. (\ref{phieqmM1}) has the following solution
\be
\phi = \pm \eta \tanh\left(\frac{z}{{\sqrt 2}R}\right)\,,
\ee
with
\be
R=\lambda^{-1/2} \eta^{-1}\,.
\ee
The rest mass of the particle is given by
\bq
m&=&\int_{- \infty}^{\infty} \rho dz = 2 \int_{- \infty}^{\infty} V dz = \frac{8 {\sqrt 2}}{3} V_{max} R  = \nonumber \\
&=& \frac{2{\sqrt 2}}{3}\lambda^{1/2}  \eta^3\,,
\eq
where $V_{max} \equiv V(\phi=0) = \lambda \eta^4/4$. Here we have taken into account that Eq. (\ref{KeqU}) implies that in the static case the total energy density is equal to $2V$. On the other hand, from Eqs. (\ref{Lagphi}) and (\ref{Ttrace}), one also has that
\be
{\mathcal L}_m=T\,, \label{LT}
\ee
where this equality is independent of the reference frame, and, consequently, it does not depend on whether the particle is moving or at rest. Also note that this result also applies to collections of particles and, in particular, to one which can be described as perfect fluid. However, unlike the result obtained for a homogeneous scalar field described by a matter Lagrangian of the form ${\mathcal L}_m(\phi,X)$, according to which the on-shell Lagrangian of a perfect fluid with proper pressure $p=0$ is ${\mathcal L}_m = 0$ (independently of its proper density $\rho$), one finds that a perfect fluid with $p=0$ made of static solitonic particles would have an on-shell Lagrangian given by ${\mathcal L}_m = T =  -\rho$. This is an explicit demonstration that the  Lagrangian of a perfect fluid depends on microscopic properties of the fluid not specified by its energy-momentum tensor.

\subsubsection{FRW spacetime}
	
Consider a $1+1$ dimensional FRW space-time with line element $ds^2=-dt^2+a^2(t) dq_z^2$, where $q_z$ is the comoving spatial coordinate and $a(t)$ is the scale factor. Taking into account that 
\be
{ \phi^{,\mu}}_{,\mu} = \left(- \Gamma^\mu_{\mu \nu} +  [\ln (f_{,\mathcal L})]_{,R} \nabla_\nu R + [\ln (f_{,\mathcal L})]_{,\mathcal L} \nabla_\nu \mathcal L \right)   \phi^{,\nu}\,,
\ee
and assuming that 
\be
f(R)=f_1(R) + {\mathcal L}_m f_2(R)\,,
\ee 
one obtains
\be
{\ddot \phi}  +   \left(H + \frac{{\dot f}_2}{f_2}\right){\dot \phi} - \nabla^2 \phi= -\frac{dV}{d\phi}\,, \label{dynphi}
\ee
where $H \equiv {\dot a} / a$ is the Hubble parameter and $\nabla^2 \equiv  d^2 /dz^2 = a^{-2} d^2 / d q_z^2$ is the physical Laplacian.

The dynamics of $p$-branes in $N+1$-dimensional FRW universes has been studied in detail in \cite{Sousa:2011ew,Sousa:2011iu} (see also \cite{Avelino:2015kdn}). There, it has been shown that the equation for the velocity $v$ of a $0$-brane in a $1+1$-dimensional FRW spacetime implied by Eq. (\ref{dynphi}) is given by
\be
{\dot v} +\left( H + \frac{{\dot f}_2}{f_2} \right) (1-v^2) v =0 \,.
\ee
Hence, the momentum of a particle in a $1+1$ dimensional FRW universe evolves as
\be
m \gamma v \propto (a f_2)^{-1}\,, \label{momev}
\ee
where $\gamma \equiv (1-v^2)^{-1/2}$.

The same conclusion could also be attained determining the macroscopic average of the microscopic energy-momentum tensor (by computing the average over a comoving volume centered at each point and containing many particles), and considering the time component of Eq. (\ref{Tcons}),
\be
{\dot \rho} +  \frac{\dot a}{a} (\rho+p)  = - ({\mathcal L}_m  + \rho) \frac{\dot f_2}{f_2} = -p \frac{\dot f_2}{f_2} \,,\label{Tcons1}
\ee
together with the assumption that the number of particles is conserved (that is $\rho \propto \gamma a^{-1}$) and that $p=\rho  v^2$. Here, $\rho = - {T^{0}}_0$ represents the proper density of the fluid, and ${\mathcal L}_m = T = -\rho + p$, where $p$ is the proper pressure of the fluid. In the following section we shall demonstrate that Eq. (\ref{momev}) also holds in $3+1$ dimensions.

\section{Particles in 3+1 dimensions}

In \cite{Avelino:2018qgt} (see also \cite{Avelino:2010bu}) it has been shown, using a Derrick-like argument, that the volume average of the on-shell matter Lagrangian of a fluid, composed of solitonic particles of fixed mass and structure, is equal to the volume average of the trace of its energy momentum-tensor, regardless of the particle's structure and constitution. Here, we shall provide an alternative derivation of this result. 

Consider a fluid composed of a statistically homogeneous and isotropic distribution of frozen solitonic particles in the comoving cosmological frame. In this case, if the particle number and mass is conserved, the energy density of the fluid evolves as $\rho \propto a^{-3}$. In order to ensure that this is always verified, one must have $S^\nu=0$ in Eq. (\ref{Tcons}) (implying that $g^{00} {\mathcal L}_m - T^{00}$ = 0) and $p=0$. These two conditions in turn imply that ${\mathcal L}_m = T$. Since this is a scalar identity, and thus independent of the reference frame, it should be independent of whether the particles are moving or at rest in the comoving cosmological frame.

In order to investigate the evolution of the momentum of a particle in a $3+1$ dimensional FRW universe, one may again consider the time component of Eq. (\ref{Tcons}), which reads
\be
{\dot \rho} +  3 \frac{\dot a}{a} (\rho +p) =  -({\mathcal L}_m  + \rho) \frac{\dot f_2}{f_2} = - 3 p \frac{\dot f_2}{f_2} \,,\label{Tcons2}
\ee
where we have taken into account that in $3+1$ dimensions ${\mathcal L}_m = T = -\rho + 3p$. Again, assuming that the number of particles is conserved (that is $\rho \propto \gamma a^{-3}$, in $3+1$ dimensions) and that $p=\rho   v^2/3$ is the proper pressure of the fluid (the factor of $3$ being associated to the number of spatial dimensions) one finds that the evolution of the momentum of a particle is still given by 
\be
m \gamma v \propto (a f_2)^{-1}\,, \label{momev2}
\ee
in a $3+1$ dimensional FRW universe. This result is identical to the one obtained in Eq. (\ref{momev}) by considering the dynamics of a solitonic particle in $1+1$ dimensions.

It is interesting to note that in the $m \to 0$, $\gamma \to \infty$ limit, relevant for photons, one obtains that ${\mathcal L}_m=T=0$. This is in agreement with the result that the electromagnetic Lagrangian,
\be
{ \mathcal L }_{EM} = -\frac{1}{4} F^{\mu \nu} F_{\mu \nu}\,,
\ee
vanishes in the case of radiation for which the relation $|{\vec E}|=|{\vec B}|$ always holds (here ${\vec E}$ and ${\vec B}$ are the electric and magnetic fields, respectively), and at odds with the assumption commonly used in the literature ${\mathcal L}_m=-\rho$.

\section{Cosmological consequences}

Let us start this section by assuming that the cosmic microwave background (CMB) has a perfect black body spectrum with temperature $T_{\rm dec}$ at the time of decoupling between baryons and photons (neglecting tiny temperature fluctuations of $1$ part in $10^5$). The spectral energy density and number density of a perfect black body are given respectively by
\be
u(\nu)=\frac{8 \pi h \nu^3}{e^{h \nu/(k_B T)}-1}\,, \qquad n(\nu) = \frac{u(\nu)}{h \nu}\,,
\ee
where $h$ and $k_B$ are, respectively, the Planck's and Boltzmann's constants, $T$ is the temperature and $E_\gamma = h \nu$ is the energy of a photon of frequency $\nu$.  In the standard scenario, assuming that the universe becomes transparent for $T < T_{\rm dec}$, the CMB radiation retains a black body spectrum after decoupling. This happens because the photon number density evolves as $n_\gamma \propto a^{-3}$ (assuming that the number of photons is conserved) while their frequency is inversely proportional to the scale factor $a$, so that $\nu \propto a^{-1}$. In the case studied in the present paper, the number of CMB photons is still assumed to be conserved (so that $n_\gamma \propto a^{-3}$) but their energy is no longer proportional to $a^{-1}$ (in our case $\nu \propto (a f_2)^{-1}$). Hence, taking into account that $n_\gamma \propto a^{-3} \propto (f_2)^3 \times (af_2)^{-3}$, the spectral energy density at a given  redshift $z=1/a-1$ after decoupling may be written as
\be
u(\nu)_{[z]} =  \frac{(f_{2[z]})^3}{(f_{2[z_{\rm dec}]})^3} \frac{8 \pi h \nu^3}{e^{h \nu/(k_B T_{[z]})}-1}\,, 
\ee
where 
\be
T_{[z]} \equiv T_{ [z_{\rm dec}]} \frac{(1+z) f_{2 [z_{dec}]}}{(1+z_{dec})f_{2 [z]}}\,.
\ee
This spectral density is similar to that of a perfect black body, except for the different normalization (we shall denote this type of spectral distortions, modifying the normalization of the spectral density, as $n$-type distortions). Also note that a small fractional variation $\Delta f_2/f_2$ on the value of $f_2$ produces a fractional change in the normalization of the spectral density equal to $3 \Delta f_2/f_2$.  

FIRAS (Far InfraRed Absolute Spectrophotometer) on board the COBE (COsmic Background Explorer) measured the spectral energy of the nearly perfect CMB black body spectrum \cite{Fixsen:1996nj,Fixsen:2009ug}. The weighted root mean square deviation between the observed CMB spectral radiance and the blackbody spectrum fit was found to be less that $5$ parts in $10^5$ of the peak brightness. Hence, we estimate that $f_2$ can vary by at most by a few parts in $10^5$ from the time of decoupling up to the present time. This provides a stringent constraint on $f(R,{\mathcal L}_m)$ theories of gravity, independently of any further considerations about the impact of such theories on the background evolution of the Universe.

\section{Conclusions}\label{conc}

In this work we have provided further evidence that the on-shell matter Lagrangian of a perfect fluid is strongly dependent on the microscopic properties of the fluid not specified by its energy-momentum tensor, giving specific examples of perfect fluids with the same energy-momentum tensor but different on-shell Lagrangians. We have shown that the on-shell matter Lagrangian of a perfect fluid, modelled as a collection of moving solitonic particles, is equal to ${\mathcal L}_m=T$, a result which is crucial for the accurate computation of the astrophysical and cosmological consequences of NMC theories of gravity where the matter Lagrangian appears explicitly in the equations of motion of the gravitational and matter fields. We have determined the modifications to the dynamics of the particles which result from the non-minimal coupling to the matter fields, and we have shown that these may lead to observable imprints on the spectral energy density of the CMB. We have imposed stringent constraints on $f(R,{\mathcal L}_m)$ gravity by requiring that the predicted spectral radiance of the CMB is consistent with the COBE-FIRAS measurement.\\

P.P.A. thanks Lara Sousa for many enlightening discussions. R.P.L.A. was supported by the Funda{\c c}\~ao para a Ci\^encia e a Tecnologia (FCT, Portugal) Grant No. SFRH/BD/132546/2017. Funding of this work has also been provided by the FCT grant UID/FIS/04434/2013. This paper benefited from the participation of the authors on the COST action CA15117 (CANTATA), supported by COST (European Cooperation in Science and Technology). 

\bibliography{Lagrangian}
\end{document}